\documentclass[doublecol,figures]{epl2} 
\usepackage{amsmath}
\usepackage{amssymb}
\usepackage{amsfonts}

\newcommand{\bv}[1]{{\boldsymbol #1}}

\makeatletter

\title{Morphology transition at depinning in a solvable model of interface growth in a random medium}
\shorttitle{Morphology transitions at depinning in a solvable model} 
\author{Hiroki Ohta\inst{1} \and Martin Luc Rosinberg\inst{2} \and Gilles Tarjus\inst{2}}
\shortauthor{H. Ohta, M. L. Rosinberg, and G. Tarjus}

\institute{                    
  \inst{1} LPTMS, CNRS-UMR  8626 and Universit\'e Paris-Sud, 91405 Orsay Cedex, France\\
  \inst{2} LPTMC, CNRS-UMR 7600, Universit\'e Pierre et Marie Curie, 
4 place Jussieu,  Paris Cedex 05, France
  }


\abstract{We propose a simple, exactly solvable, model of interface growth in a random medium that is a variant of the zero-temperature 
random-field Ising model on the Cayley tree. This model is shown to have a phase diagram (critical depinning field versus disorder strength) 
qualitatively similar to that obtained numerically on the cubic lattice. We then introduce a specifically tailored random graph that allows 
an exact asymptotic analysis of the height and width of the interface. We characterize the change of morphology of the interface 
as a function of the disorder strength, a change that is found to take place at a multicritical point along the depinning-transition line. }

\begin{document}
\maketitle

\section{Introduction}

The motion of a driven interface in a random medium keeps attracting wide attention in condensed-matter and statistical physics, 
as it occurs in many different physical processes such as fluid invasion in porous media, domain wall motion in magnetic systems, contact 
line motion in wetting, etc... \cite{K1998,F1998}.
In such systems, the driven interface displays a transition (at zero temperature) from a pinned phase to a moving phase as the driving force 
surpasses some finite threshold value. This depinning transition has been extensively studied numerically and theoretically, in particular 
for continuum models of elastic interfaces in the presence of quenched disorder \cite{CGL2000,FBKR2013}. 

Driven interfaces also display remarkable changes in the domain growth morphology as a function of the disorder strength, an issue that 
has been explored in detail within the random-field Ising model (RFIM) for various dimensions, coordination numbers, and distributions of 
the random fields \cite{JR1991,NKMR1993,NU1998,KR2000,KR2010}. Specifically, in three dimensions, for an unbounded distribution 
of the random fields (e.g. Gaussian),  one observes a transition from a compact-growth regime with a self-affine interface at low disorder 
to a self-similar, percolation-like growth regime at high disorder \cite{KR2000}. These two regimes are separated by a multicritical point 
and described by  different sets of critical exponents.  

The RFIM on a three-dimensional lattice, however, can only be studied numerically, and it would be  useful to have a simple and exactly 
solvable model of interface growth that exhibits, at least qualitatively, the same type of morphology changes. This goal is achieved in the 
present Letter where we introduce an interface-growth dynamics for the RFIM on a Cayley tree for which  evolution equations can be 
written exactly, and solved, in the thermodynamic limit \cite{note0}. We show that the phase diagram of the model (critical depinning field vs. 
disorder strength) is qualitatively similar to that obtained on the cubic lattice \cite{KR2000}. As there is no clear-cut definition of the 
height and width of the  interface on a Cayley tree, we then introduce a specifically tailored random graph (randomly connected chains) 
with a local tree structure that still makes the problem analytically tractable. The phase diagram is the same as for the Cayley tree 
and is shown to be indeed associated with the two distinct growth morphologies observed on the cubic lattice \cite{KR2000}.

\section{Model and dynamics}

We consider the RFIM on a Cayley tree of degree (coordination number) $c+1$. 
The Hamiltonian is given by
\begin{equation}
{\mathcal H}(\bv{\sigma}) = -\frac{1}{2}
\sum_{i}\sum_{j\in \partial i} \sigma_i\sigma_j -\sum_i (H + h_i)\sigma_i,
\end{equation}
where $i$ runs over all sites (nodes) of the graph, $\bv{\sigma}\equiv\{\sigma_i\}$ with $\sigma_i=\pm 1$, 
$\partial i$ is the set 
of sites that are directly connected to $i$, $H$ is the external field, 
and the random fields $\{h_i\}$ are independently taken from a Gaussian distribution $\rho(h)$ with zero mean and variance $R^2$. 
The strength of the exchange interaction is taken as the energy unit.

The Cayley tree can be visualized as a root (origin) from which  different generations emanate. At the $n$th generation level then, 
there are  $c^n$ sites. To describe interfaces, however, it is more convenient to start from an outer boundary at a very large 
generation level $L$ and to proceed inward towards the origin: we therefore label a site $i$ by two indices, $p$ and $z$, with $z$ 
denoting the number of generations counted from the outer boundary ($z=0$ at the boundary and $z=L$ at the origin) and $p$ 
labeling the $c^{L-z}$ sites at the level $z$. 

We study an interface-growth dynamics that is analogous to the zero-temperature front-propagation dynamics 
considered in \cite{JR1991,NKMR1993,NU1998,KR2000,KR2010}. The interface is defined as the collection of spins that  
are  -1 (``down'') and have at least one direct neighbor on the graph that is $+1$ (``up''). (Note that this includes spins on the 
foremost  ``front'' as well as at the interfaces and isolated down spins left behind.) The initial condition at time $t=0$ corresponds 
to all spins at the boundary $z=0$ in the up state and all the other spins in the down state. The configuration of spins 
is then evolved at a fixed external field $H$ with the following rule:  only spins \textit{at the interface} are allowed to flip and a spin 
at site $i$ on the interface at time $t-1$ flips to $+1$ at time $t$ when the local field $H+h_i +\sum_{j\in \partial i}\sigma_j(t-1)$ is positive 
(which then lowers the total energy  of the system); all unstable spins at the interface are flipped simultaneously 
at each time step (parallel dynamics). It is worth stressing that this interface growth dynamics for the RFIM at zero temperature 
is different than the dynamics used to describe the hysteresis loop of the magnetization in the same model \cite{SDKKRS1993}.

\section{Depinning transition}

For a sufficiently large negative value of $H$, the interface remains pinned whereas for a sufficiently large positive value, 
it keeps moving. In between  a depinning transition takes place at a critical value $H_c$ that depends on the disorder 
strength $R$. This phenomenon can be captured by studying the probability $P_t$ that 
a randomly chosen spin at time $t$ and at level $z=t$ is up: in the large $t$ limit, $P_t=0$ in the pinned 
phase and $P_t>0$ in the moving one. (By construction, the interface cannot be higher than level $t$ at time $t$.) Thanks to the tree structure of the graph, 
one can easily write down a recursion equation for $P_t$:
\begin{equation}
\label{eqPt1}
P_t=\sum_{k=1}^{c} {\binom ck} p_k (H) P_{t-1}^k[1-P_{t-1}]^{c-k},
\end{equation}
where the $k$th term of the above sum is the probability that the chosen site at level $t$, which 
has a down neighbor at level $t+1$, has $k\geq 1$ up and $c-k$ down neighbors at level $t-1$ and has a positive local field;  
$p_k(H)$ is then the probability that the local field is positive when $k$ neighbors are up and $c+1-k$ are down, \textit{i.e.}  
$p_k(H)= \int_{- H -2k + (c+1)}^{\infty} dh \rho(h)$.

For large $t$, $P_t$ tends to a fixed point $P^*$ given by the self-consistent equation obtained from Eq. (\ref{eqPt1}).
This is a polynomial equation that has generically $c$ solutions, including $P^*=0$. For concreteness we discuss 
more specifically the case $c=3$, but similar results are obtained for $c>3$. 
The two nontrivial solutions are then given by $P_{\pm}^*(H,R)=-(v \pm \sqrt{v^2-4uw})/2u$,
where $u\equiv (p_3-3p_2+3p_1)$, $v\equiv 3(p_2-2p_1)$, and $w\equiv (3p_1-1)$.

As one increases $H$ from a large negative value, the only physical fixed point is at first $P^*=0$ (the two other fixed points 
are initially real but unphysical, and then become complex).  Below some critical disorder $R_{\rm c}$, the two complex solutions  
merge ($v^2=4uw$) at a critical field $H=H_{\rm sn}(R)$  and become real: $P_+^*=P_-^*=-v/(2u)$, with $0<-v/(2u)\leq 1$. $P_+^*$  
is then the stable fixed point for $H\ge H_{\rm sn}(R)$ while $P_-^*$ remains unstable. This corresponds to a \textit{saddle-node bifurcation}. 
On the other hand, for a strong disorder $R>R_{\rm c}$,  $P_+^*$ becomes the physical, stable fixed point (while $P_-^*$ is unphysical) 
by passing through zero from the negative side ($w=0$) at a critical field $H=H_{\rm tc}( R)$. This is a  \textit{transcritical bifurcation}.  
The two distinct bifurcations meet at a multicritical point $(R_{\rm c},H_{\rm c}(R_{\rm c}))$ where $P_+^*=P_-^*=0$ 
(implying $v=w=0$, {\it i.e.}, $p_2=2p_1=2/3$): this yields $R_{\rm c}= 2.32165\cdots$ and $H_{\rm c}(R_{\rm c})=1$.  

One can also describe the long-time dynamical behavior of $P_t$ in the thermodynamic limit $L\to\infty$.
Below the line formed by $H_{\rm sn}(R) $ and $H_{\rm tc}(R)$, $P_t$ approaches zero exponentially fast 
as $t\rightarrow \infty$ and one is in the pinned phase. Above 
the line, $P_t$ approaches the nonzero fixed-point value $P_+^*(H,R)$ exponentially fast and one is  
in the moving phase. The line corresponds to a depinning transition whose nature is different above and below $R_{\rm c}$:

(i) Above $R_{\rm c}$, the transition is continuous and $P_t$  approaches zero with a power law $t^{-1}$. When $H$ is close to $H_{\rm tc}(R)$,  
$P_t$ can be cast in a scaling form
\begin{equation}
\label{scaling1}
P_t\simeq \vert H- H_{\rm tc}\vert f_{\pm}(\vert H- H_{\rm tc}\vert t)
\end{equation} 
with $f_-(-\infty)=0$, $f_+(\infty)>0$, and $f_{\pm}(x\rightarrow 0)\sim x^{-1}$.  
(More precisely, $P_t\simeq (-1/v)t^{-1}$ with $v<0$ for $H=H_{\rm tc}$.)  
The transition then belongs to the same universality class as the mean-field percolation.
 
(ii) Below $R_{\rm c}$, the transition takes place with a jump in the asymptotic value of $P_t$. On the moving side, near the transition, 
there is a slow approach to the asymptotic value $P_+^*(H_{\rm sn},R)=-v/(2u)$ 
and $P_t$ can be cast in a scaling form
\begin{equation}
\label{scaling2}
P_t-P_+^*\simeq (H- H_{\rm sn})^{1/2}f_{+}( (H- H_{\rm sn})^{1/2} t)
\end{equation} 
with $f_+(\infty)>0$ and $f_{+}(x\rightarrow 0)\sim x^{-1}$. (More precisely, $P_t-P_+^*\simeq (2/v)t^{-1}$ with 
$v>0$ for $H=H_{\rm sn}$ \cite{note2}.)
On the pinned side, close to the transition, the trajectory of $P_t$ is first attracted to the fixed point $P_+^*(H_{\rm sn},R)$ as $t$ 
increases, but eventually, at a rather well defined crossover time $t_{\rm co}$, $P_t$ goes exponentially to the stable fixed point $0$. 
The crossover time $t_{\rm co}$ diverges as $(H_{\rm sn}-H)^{-1/2}$ as $H \nearrow H_{\rm sn}$. 
Thus, the transition at the saddle-node bifurcation has a mixed, first- and second-order, character 
with a jump in the order parameter and a diverging characteristic time scale. 
A similar situation is found in the zero-temperature Glauber dynamics of the random-field Ising model \cite{OS2010}, the for $k$-core 
(or bootstrap) percolation \cite{CLR1979,SLC2006,IS2009} and in kinetically constrained models \cite{SBT2005}.

(iii) Finally, near the critical point $R_{\rm c}$ separating the two types of transitions, one finds that along the saddle-node transition line, 
$P^{*}_+(H_{\rm sn},R) \sim (R_{\rm c}-R)$ when $R \nearrow R_{\rm c}$.
Exactly at $R_c$, $P_t$ can also be put in a scaling form 
\begin{equation}
\label{scaling3}
P_t\simeq \vert H-H_{\rm c}\vert ^{1/3}f_{\pm}(\vert H-H_{\rm c}\vert^{2/3}t)
\end{equation} 
with $f_-(-\infty)=0$, $f_+(\infty)>0$, $f_{\pm}(x\rightarrow 0)\sim x^{-1/2}$.  
(More precisely, $P_t\simeq(t^{-1/2}+t^{-3/2}/4)/\sqrt{-2u}$ with $u<0$ at $(R_{\rm c},H_{\rm c})$.)

\begin{figure}
\begin{center}
\includegraphics[width=7.0cm,trim=10 1 1 6,clip]{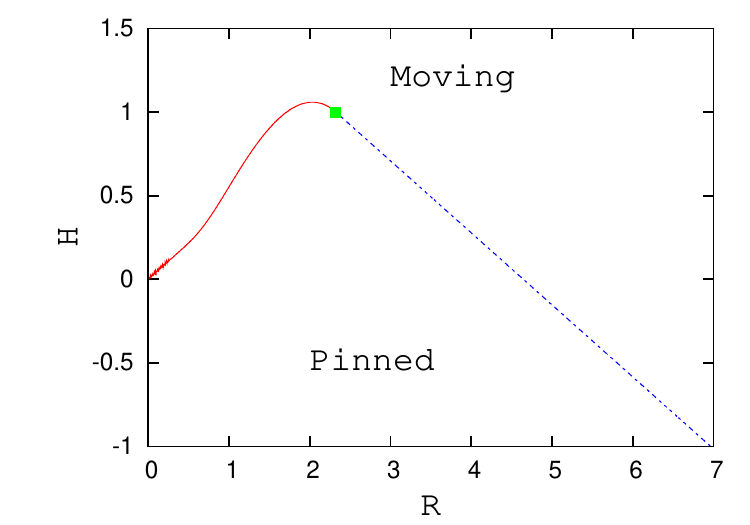}
%\raisebox{0.45cm}
\caption{(Color online) Phase diagram of the interface-growth model on a Cayley tree with $c=3$: Critical field $H_{\rm c}$ 
versus disorder strength $R$. The depinning transition changes from a saddle-node bifurcation (red line, $R<R_{\rm c}$), 
to a transcritical bifurcation (blue line, $R>R_{\rm c}$), through the multicritical point (green point, $H_{\rm c}=1$, 
$R_{\rm c}\simeq 2.32165$).}
\label{phase}
\end{center}
\end{figure}

The phase diagram corresponding to the interface growth model for the RFIM on a Cayley tree with $c=3$ is shown in Fig.~\ref{phase}.
There is a striking resemblance with the phase diagram obtained for the same model  on the cubic lattice: see Fig.~3 of \cite{KR2000}. 
In both cases, the nature of the depinning transition changes at a critical value of the disorder close to the maximum of the transition 
line $H_{\rm c}(R)$. On a cubic lattice the growth proceeds with a self-affine interface at low disorder and with a self-similar 
one at high disorder. On the Cayley tree, a similar pattern is found with a continuous, percolation-like, transition at high disorder and a 
mixed transition with features akin to those of the self-affine case at low disorder (see also below). In addition, just as observed 
on Euclidean lattices, only the continuous percolation-like transition survives when the coordination number is small enough 
(here, $c=2$, which can be compared to $d=2$). Furthermore, all transitions disappear when $c=1$, which corresponds to $d=1$.

\section{Interface morphology on randomly connected chains} 

The distinct signature of the changes of growth mechanism at the depinning transition on a cubic lattice lies in the morphology of 
the interface. A central quantity to characterize the latter is the height of the ``front'' (the foremost part of the interface) at 
time $t$, $l_p(t)$, for each position $p=1,\cdots,c^L$ on the initial boundary; $l_p$ can be defined as the maximum $z$ at which one finds an up spin 
in the shortest path connecting the tree origin ($z=L$) to the position $p$ at the boundary ($z=0$)\cite{note1}.
However, there are some difficulties to consider this quantity on the Cayley tree, due to the peculiar, 
hyperbolic-like, geometry of the latter, with an exponentially decreasing number of sites with increasing 
height $z$. It is hard to visualize and assess the interface morphology in such a case, and numerical experiments 
are extremely demanding.

\begin{figure}
\begin{center}
\includegraphics[width=5cm,trim=1 1 1 1,clip]{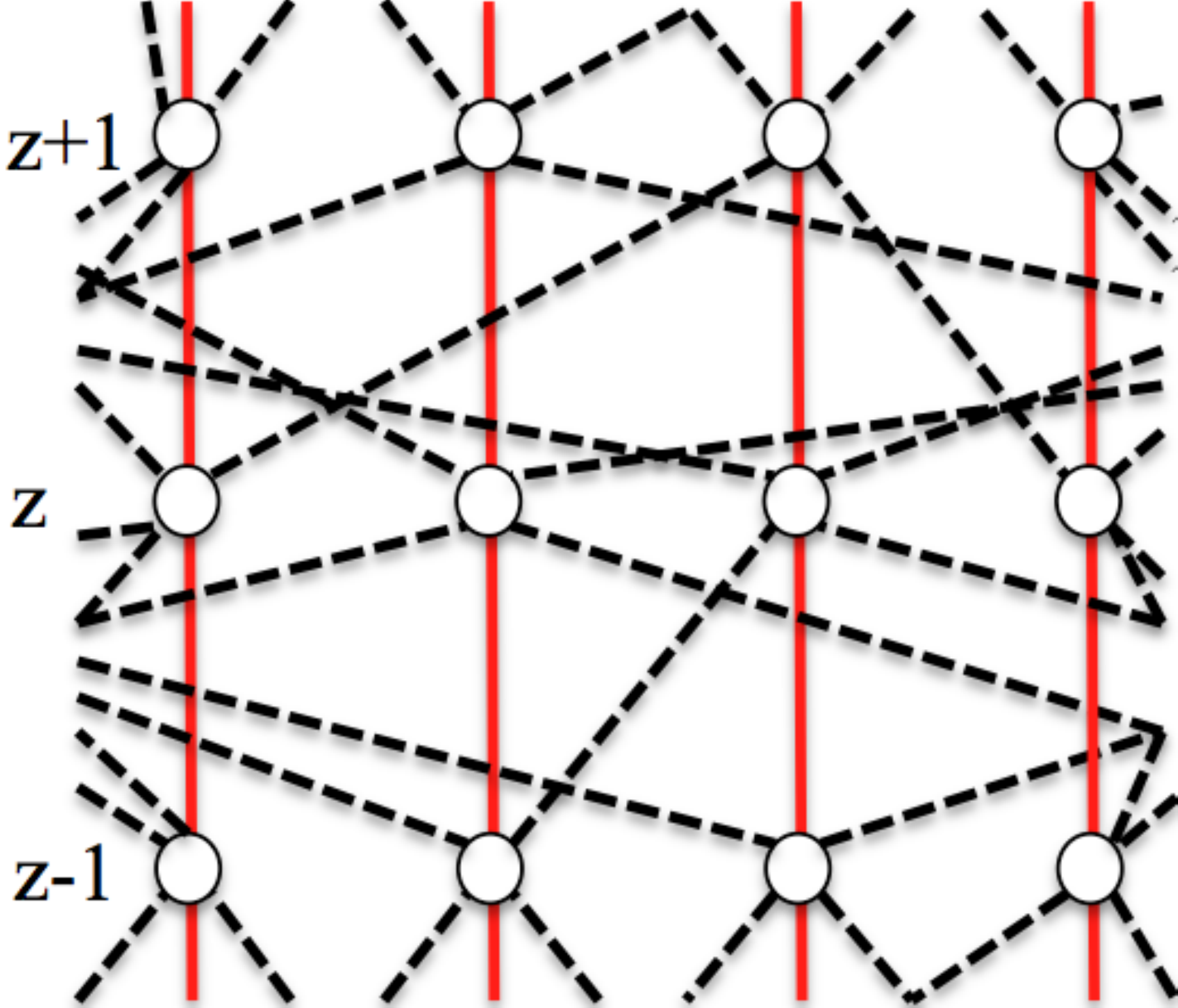}
\caption{(Color online) Illustration of the randomly connected chain graph with $b=2$. 
Each site at height $z$ is connected to its two nearest neighbors at height $z-1$ and $z+1$ along the same chain (vertical red lines) 
and randomly connected to $b$ additional sites at level $z-1$ and to $b$ additional sites at level $z+1$ 
on the other chains (dotted lines).}
\label{rcc}
\end{center}
\end{figure}

To sidestep this issue and provide some insight on possible changes in the interface morphology 
while retaining the  analytically solvable character of tree-like graphs, we introduce a specifically tailored random graph, 
which can be described as randomly connected chains (see Fig.~\ref{rcc}). It consists of $L_{x}$ parallel vertical chains of connected 
sites of height $L \geq L_x$. A site at height $z$ is (deterministically) connected to its two nearest neighbors 
on the same chain, as well as randomly connected to $b$ additional sites at level $z-1$ and to $b$ additional sites  
at level $z+1$ on the other chains. The total coordination number of each site is therefore $2(b+1)$, except for the 
bottom sites forming the initial interface (at $z=0$) 
and the top sites at $z=L$, for which it is $b+1$.

In spite of the presence of chains, the graph has a local tree structure. From heuristic arguments one expects that the 
typical loop length essentially behaves as $\log L_x$ when $L_{x}\rightarrow \infty$, so that loops become irrelevant in 
the thermodynamic limit. A rough argument goes indeed along the usual lines \cite{Ballobas,MezardMontanari} as follows. 
Consider a random site, say at height $z$, and assume that in a ball of large radius $\ell$ 
the graph is a tree: the number of sites in the ball then goes essentially as $[2(b+1)]^{\ell}$. 
On the other hand, the total number of sites in the strip of the  graph including the ball is $2\ell\, L_x$. The tree-like  
assumption then breaks down and loops must appear when $[2(b+1)]^{\ell}\sim 2\ell\, L_x$, \textit{i.e.} when $\ell \sim \log L_x$ in the 
large $L_x$ limit. A somewhat refined argument accounting more precisely for the presence of (determistic) chains of links is given in 
the online supplementary material\cite{SupMat}, but it leads to the same conclusion, up to subdominant terms. (Although the height $L$ does not 
explicitly appear in the reasoning, it should of course be large, which is satisfied by requiring that $L$ is of the order of $L_x$.)  Note finally that the tree-like character of the graph in the limit $L_x \rightarrow \infty$  is also confirmed by the 
excellent agreement between numerical simulations and theoretical expressions (see below and online supplementary information\cite{SupMat}).

\begin{figure}
\begin{center}
\includegraphics[width=4.41cm]{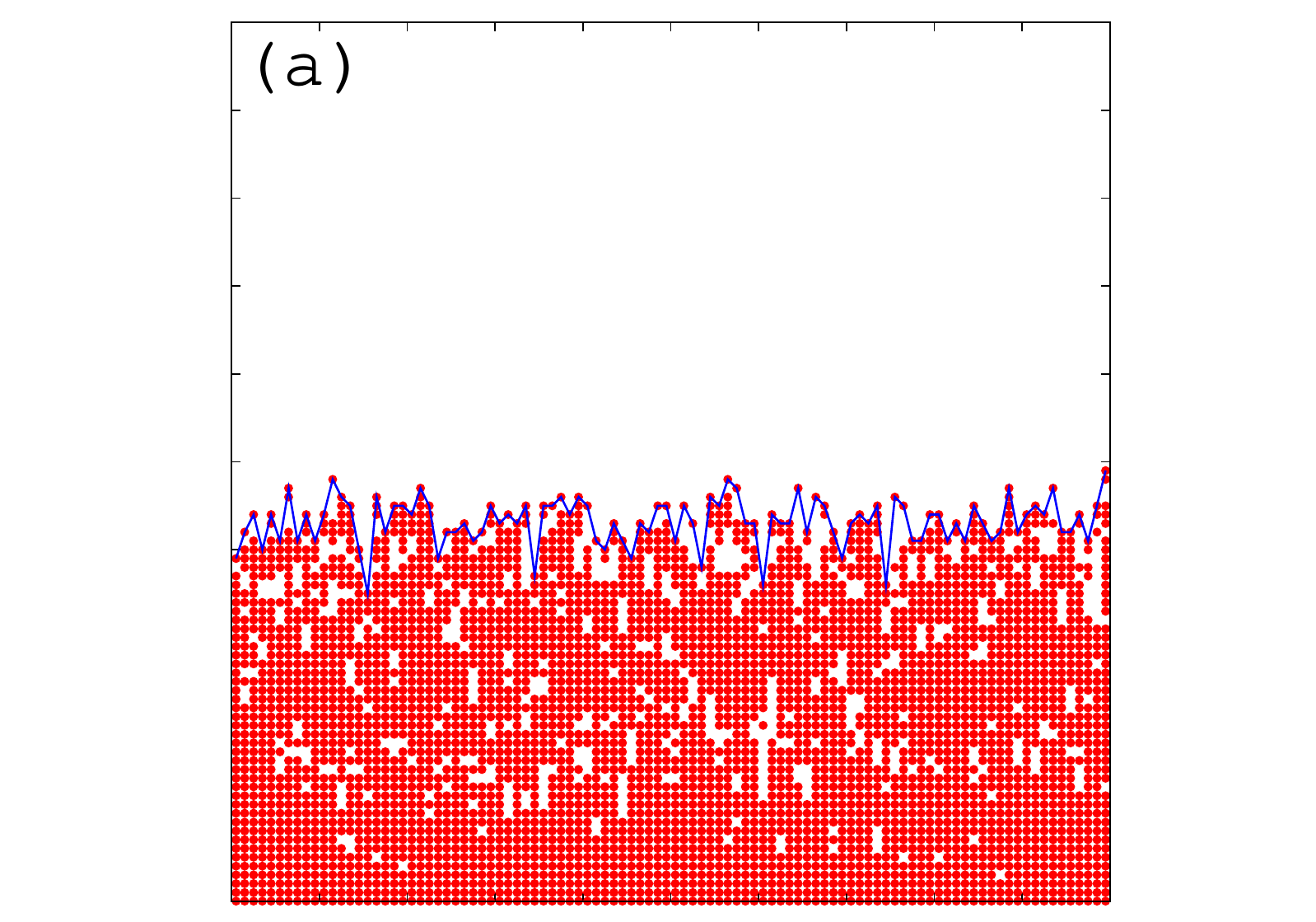}\includegraphics[width=4.41cm]{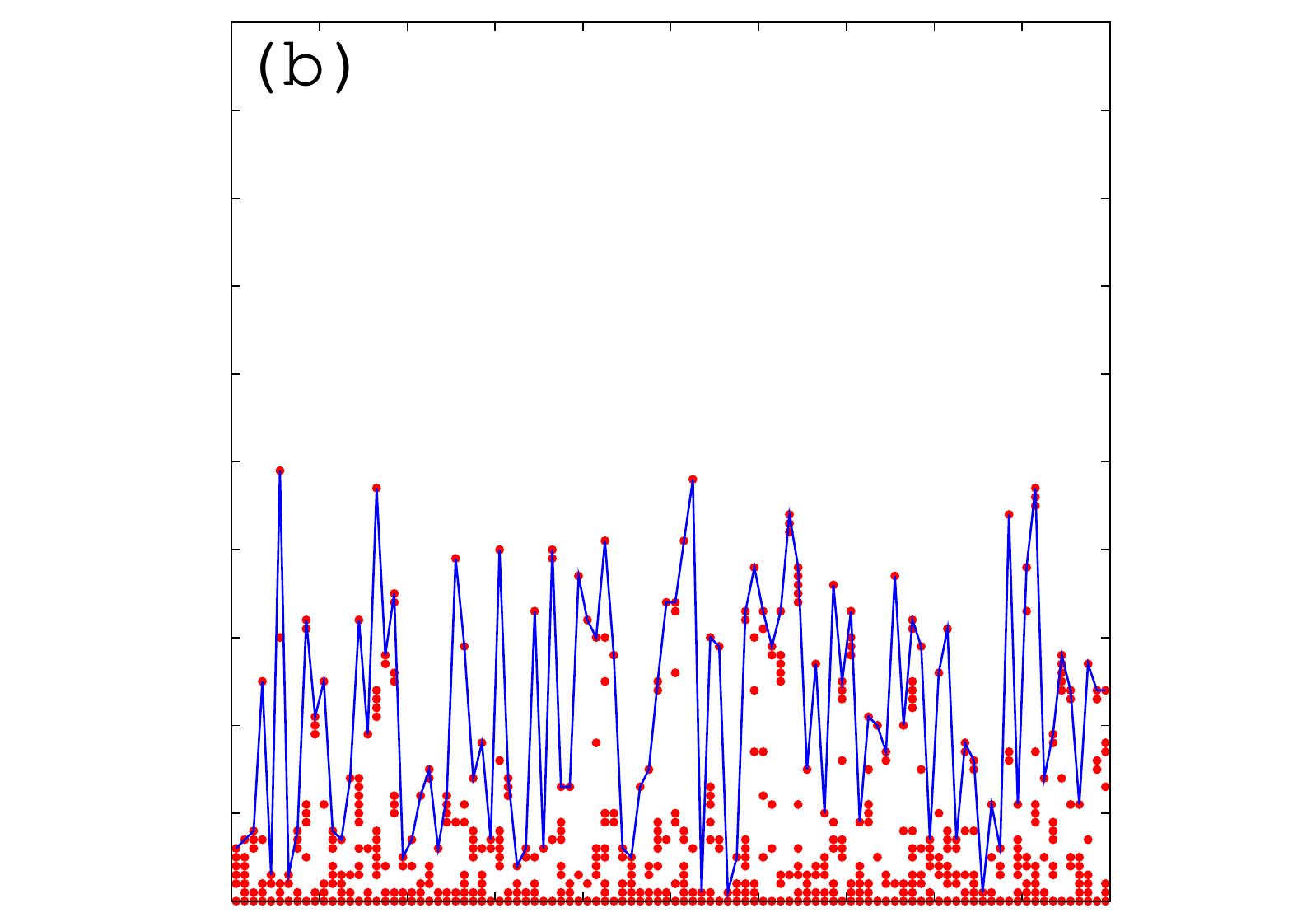}
%\centerline{\includegraphics{rcc.pdf}}
%\centerline{\includegraphics{picR6H142.pdf}}
\caption{(Color online) 
Spin configurations near the depinning transitions on randomly connected chains with $b=2$ 
(up spins are shown as red dots and the front $\{l_p\}_{p=1}^{L_{x}}$ is shown as a blue line) with $L_{x}=L=100$ and $t=50$: 
(a) low-disorder regime, $R=1, H=2.54$. 
(b) high-disorder regime, $R=6, H=1.42$.}
\label{inter}
\end{center}
\end{figure}

The main advantage of the randomly connected chains is that there is now a natural definition of the front height $l_p$ by considering 
the situation on the $p$th chain with $p=1,\cdots, L_{x}$: $l_p$ corresponds to the highest up spin on the chain. The interface growth  proceeds in a similar way as on the Cayley tree (see above) and one follows the motion of the interface from the initial boundary at $z=0$. 
However, to reproduce the results previously obtained on the Cayley tree, it turns out to be necessary to introduce a modification in the dynamical rule. Otherwise, the peculiar structure of the randomly connected 
chains allows for additional back-reaction effect compared to the Cayley tree: a spin on the front that flipped at time $t$ can  influence the front on another chain at a later time by triggering the flip of spins at lower heights. To avoid this effect, spins at height $z<t$ are not allowed to flip at time $t$. This modification is the price to pay for  mimicking real-space interfaces while retaining a tree-like structure.  

With this additional constraint, 
the probability $P_t$ that a randomly chosen spin flips up at $z=t$ can still be described by an exact recursion equation (in the thermodynamic 
limit). At time $t$ all spins at $z=t+1$ are down by construction, and one only needs to consider the various possibilities 
for the $b+1$ spins at level $z=t-1$. One can then use the statistical independence of the branches reaching the chosen site at $z=t$ to derive
\begin{eqnarray}
P_t = \sum_{k=1}^{b+1} {\binom {b+1}{k}} \tilde p_k (H)\, P_{t-1}^k\,[1-P_{t-1}]^{b+1-k}, \label{eq}
\end{eqnarray}
where $\tilde p_k(H)= \int_{- H -2k + 2(b+1)}^{\infty} dh \rho(h)$. By comparison with Eq.~(\ref{eqPt1}), one 
concludes that the interface-growth dynamics on the randomly connected chains \textit{exactly} reproduces that on the Cayley tree 
provided one chooses $b=c-1$ and shifts the external field from $H$ to $H+c-2b-1=H-b$. 
For $b=2$, the phase diagram of the model  is thus identical to the one shown in Fig.~\ref{phase}, with the $y$-axis shifted by $2$.   
To illustrate the change of growth morphology as a function of disorder, we show in Figs.~\ref{inter} (a) and (b) the results of 
numerical simulations performed near the depinning transition
in the two different regimes. It is manifest that the interface roughness is much larger in the percolation-like regime at high disorder.

The interest of the model is that it remains analytically tractable. To provide further information we consider the probability 
$Q_t(l)$ that the front height $l_p(t)$ on a randomly chosen chain $p$ is equal to $l$. (Note that due to the modification of the dynamics, 
the only meaningful interface is now the foremost front as spins behind are no longer allowed to evolve.) 
A closed-form equation for $Q_t(l)$ can be derived in the thermodynamic limit thanks to the local tree structure 
of the randomly connected chains. 
It is useful to first introduce the conditional probability $q^{\pm}_t$ that a spin at a site 
$\{p,z=t\}$ flips at time $t$ provided that the spin just below at $\{p,z=t-1\}$ is equal to $\pm 1$; 
by using the statistical independence of the branches arriving at the site under consideration, 
it can be expressed as
\begin{equation}
 q^{\pm}_t=\sum_{k=1-\delta_{\sigma,1}}^{b} {\binom bk} \tilde p_{k+\delta_{\sigma,1}}(H) \, P_{t-1}^k\,[1-P_{t-1}]^{b-k},
\end{equation}
where $\delta_{\sigma,1}$ is a Kronecker symbol and $\sigma=\pm 1$. Recalling that at time $t$ spins are allowed to flip only at height $z=t$ and not below and that a spin at the front is the highest up spin along its chain, 
one immediately finds
\begin{equation}
 Q_t(l=t)=P_t. \label{eq1}
\end{equation}
The probability that the front height is at $l=t-1$ at time $t$ 
requires that a spin flipped at $z=t-1$ at time $t-1$ and that the spin 
directly above it does not flip at time $t$, hence,
\begin{equation}
Q_{t}(l=t-1)=P_{t-1}[1-q^+_t].\label{eq2}
\end{equation}
Finally, along the same lines, one easily derives that for lower heights $0\leq l \leq t-2$,
\begin{equation}
Q_t(l)=Q_{t-1}(l)[1-q^-_t].\label{eq3}
\end{equation}
From $Q_t(l)$ one can compute the moments $\langle l^n\rangle_t\equiv \sum_{l=0}^t l^n Q_t(l)$ and derive 
the mean height $l_0(t)\equiv\langle l\rangle_t$ and the mean width $W_0(t)\equiv[\langle l^2\rangle_t-l_0(t)^2]^{1/2}$. To study 
the asymptotic behavior at large time of $l_0$ and $W_0$, it is convenient to use the above equations to derive 
\begin{align}
&l_{0}(t)=tP_{t}+(t-1)P_{t-1}[1-q^+_{t}]+X_{t}^{(1)}\ , \\
&W_{0}(t)^2+l_0(t)^2=t^2P_{t}+(t-1)^2P_{t-1}[1-q^+_{t}]+X_{t}^{(2)} \ ,
\end{align}
where $X_t^{(n)}$ for $n=1,2$ satisfies the following equation:
\begin{align}
X_{t}^{(n)}=[1-q^-_{t}]\left (X_{t-1}^{(n)}+(t-2)^nP_{t-2}[1-q^+_{t-1}]\right). \label{Y0}
\end{align}

The asymptotic analysis for $t\rightarrow \infty$ then leads to the following results:

 (i) In the pinned phase, both $l_0(t)$ and $W_0(t)$ 
are of ${\cal O}(1)$, while in the moving phase, $l_0(t)\simeq t$ and $W_0(t)={\cal O}(1)$. 

(ii) At the discontinuous (saddle-node) transition for $R<R_{\rm c}$, $l_0(t)\simeq t$ and 
$W_0(t)={\cal O}(1)$. 

(iii) At the continuous percolation-like transition (transcritical bifurcation) for $R>R_{\rm c}$, $l_0(t)\simeq a t$ with $a=2/(8-9\tilde p_2)<1$ 
and $W_0(t )\simeq b t$ with $b=3(2 -3\tilde p_2)/[(8-9\tilde p_2)(7-9\tilde p_2)^{1/2}]<1$. 

\begin{figure}
\begin{center}
\includegraphics[width=4.41cm,trim=4 1.8 2.1 3,clip]{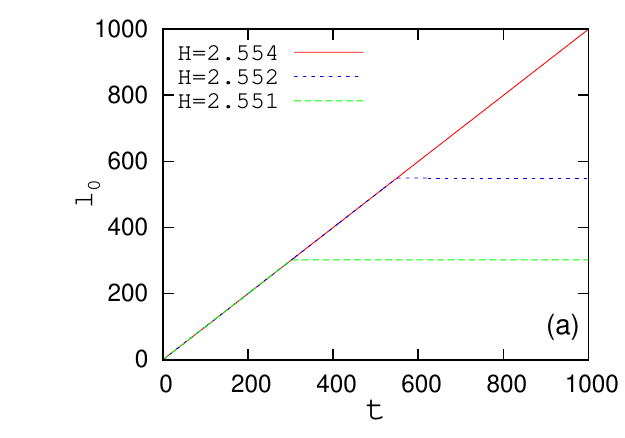}\includegraphics[width=4.41cm,trim=4 1.8 2.1 3,clip]{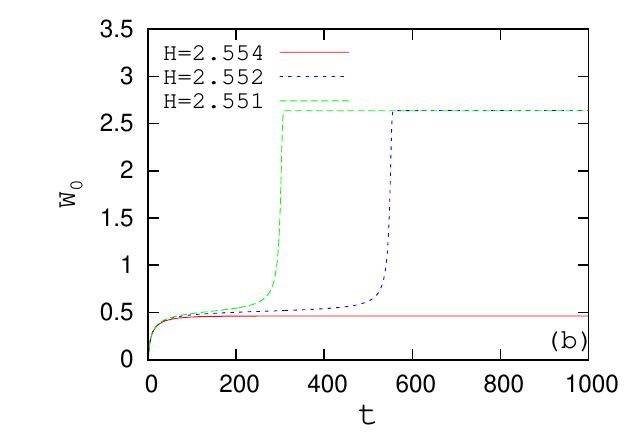}
\includegraphics[width=4.41cm,trim=4 1.8 2.1 3,clip]{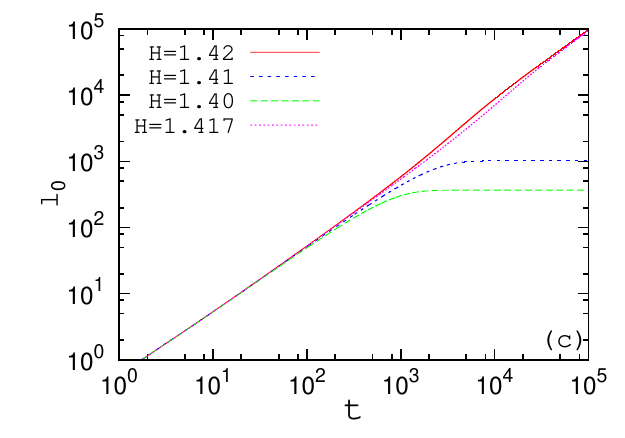}\includegraphics[width=4.41cm,trim=4 1.8 2.1 3,clip]{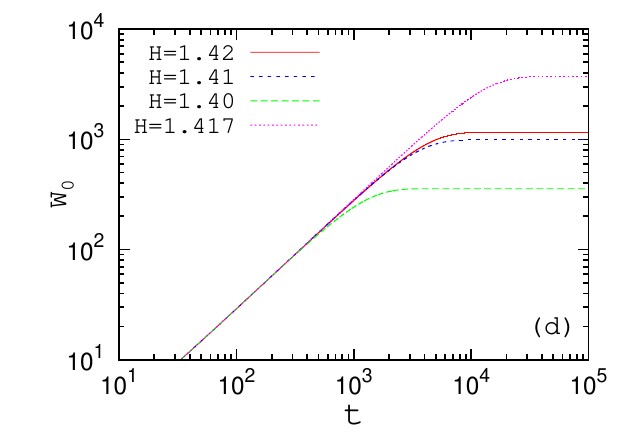}
\caption{(Color online) Mean height $l_0(t)$ and width $W_0(t)$ of the front on randomly connected chains with 
$b=2$, as obtained from Eqs.~(7)-(9).
(a) and (b): Behavior for $R<R_{\rm c}$ near the saddle-node bifurcation ($R=1$, $H_{\rm sn}(1) \simeq 2.5524$).
(c) and (d): Behavior for $R>R_{\rm c}$ near the transcritical bifurcation ($R=6$, $H_{\rm tc}(6)\simeq 1.4156$).
Note the presence of a crossover time $t_{\rm co}$ at which the behavior changes in the pinned phase and which diverges as one 
approaches the depinning transition. One can associate to it a diverging crossover length.}
\label{analytic}
\end{center}
\end{figure}

(iv) Finally, at the multicritical point for $(R_{\rm c},H_{\rm c})$, $l_0(t)\simeq  t$ and 
$W_0(t)\simeq c t^{1/2}$ with $c=3/\sqrt{-2u}\simeq 0.664$. The full solution is illustrated in Fig.~\ref{analytic}.

Thus, as anticipated by looking at Fig.~\ref{inter}, the behavior of $l_0(t)$ and $W_0(t)$ 
provides a clear signature of the change in the morphology of the front at depinning as a function of disorder, 
from rather compact for $R<R_{\rm c}$ to rough and percolation-like for $R>R_{\rm c}$.

\begin{figure}
\begin{center}
\onefigure[width=6cm,trim=8 8 5 3,clip]{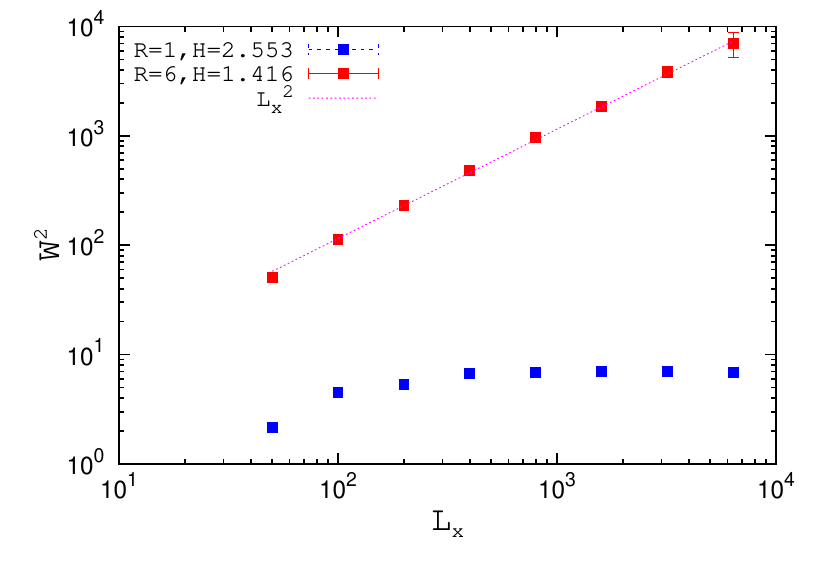}
%\centerline{\includegraphics{Wl.pdf}}
\caption{(Color online) $W(t=L_{x},L_{x})^2$ as a function of $L_{x}$ near the depinning transition on randomly connected chains 
with $b=2$. Lower (blue) curve: saddle-node bifurcation regime; Upper (red) curve: transcritical (percolation-like) bifurcation regime.}
\label{rough}
\end{center}
\end{figure}

In addition to the analytical study we have also carried out numerical simulations of the interface-growth process. 
It allows us, on the one hand, to confirm that the randomly connected chains is equivalent to a tree in the thermodynamic limit and, on the 
other hand, to assess finite-size characteristics of the interface (front) that are hard to compute analytically. We focus on the 
mean width at time $t$ for graphs of lateral extension $L_{x} \gg 1$ (typically from $50$ to $6400$):
\begin{equation}
W(t,L_{x})\equiv\overline{[\langle l_p(t)^2\rangle - \langle l_p(t)\rangle^2]}^{1/2},
\end{equation}
where the overline denotes an average over the random fields  and the random graphs (the number of samples 
varies from $4096$ for $L_x=50$ to $32$  for $L_x=6400$), 
and $\langle \; \rangle$ an average over the $L_{x}$ chains, 
$(1/L_{x})\sum_{p=1}^{L_{x}}$. When $L_{x}\rightarrow \infty$, 
$W(t,L_{x})$ goes to $W_0(t)$ already computed (See also the supplementary information\cite{SupMat}). As can be seen in Fig. \ref{rough}, 
$W(t=L_{x},L_{x})$ at the depinning transition appears to follow a power law $(L_{x})^{\alpha}$ 
at large $L_{x}$, where $\alpha$ can then be taken (with a pinch of salt due to the nature of the graph) 
as a ``roughness'' exponent. 
The numerical data leads to $\alpha=1/2$ for $R>R_{\rm c}$ and $\alpha=0$ 
for $R<R_{\rm c}$. (The data for $R=R_{\rm c}$, not shown here, seem compatible with $\alpha \simeq1/2$ but the 
convergence to the asymptotic limit is much slower so that the estimate is not as reliable as for the other cases.)
From all the above results, the mean interface width at the depinning transition 
is then expected to follow a scaling form
\begin{equation}
\label{scaling4}
W(t,L_{x})\simeq (L_{x})^{\alpha}\, w[t/(L_{x})^{\alpha/\beta}]
\end{equation}
with $w[\infty]={\cal O}(1)$, $w[y\rightarrow 0]\sim y^\beta$, $\beta=0$ for the saddle-node bifurcation 
(on the moving side), $\beta=1$ for the transcritical bifurcation, and $\beta=1/2$ for the multicritical point. 

\section{Concluding remarks}

In summary, we have proposed a solvable model of interface growth in a random medium on a Cayley tree that displays a 
morphology  transition at depinning as a function of disorder strength. Along the depinning line, a multicritical point separates a low-disorder 
regime where the transition has a mixed continuous-discontinuous character and is described by a saddle-node bifurcation and 
a high-disorder regime where the transition is continuous and characterized by a  transcritical bifurcation. 

To analyze the interface (front) morphology in more detail, we have introduced a trick: we have replaced the problem on a Cayley tree 
by one on a graph formed by randomly connected chains. With some  adjustments in the dynamics, the latter is described by the same recursion 
equations as the former, but it allows a clear definition and visualization of the front. The results then illustrate that the low-disorder 
regime corresponds to a compact growth while the high-disorder one corresponds to percolation-like growth, showing that 
the multicritical point can be associated with a morphology transition. Remarkably, the phase diagram is similar to that of the same model on the cubic lattice \cite{KR2000}. In the latter case the multicritical point was conjectured to coincide with the maximum observed in the 
depinning transition line $H_{\rm c}(R)$ (error bars in the simulations make it difficult to settle this issue) \cite{KR2000}. We show here 
analytically that on tree-like graphs this  maximum unambiguously takes place in the low-disorder region (see Fig.~\ref{phase}), 
below the multicritical disorder strength.
 
\acknowledgments
We are grateful to A. Rosso, T. Sasamoto and G. Semerjian for useful discussions and H. O. thanks the Yukawa Institute for 
Theoretical Physics where this work was initiated. This work is also supported by EC-Grant ``STAMINA'' No. 265496.

\textbf{Supplementary Information: loops in the randomly connected chain graph}
\\

We have given in the main text a rough argument showing that loops become significant when they have a length $\ell$ of the 
order of $\log L_x$, the lateral dimension of the graph, \textit{i.e.} the number of chains (see Fig. 2). Here we provide a refined, 
but still heuristic, derivation of the typical loop length (the typical length of the shortest loop through a randomly chosen point) 
in a randomly connected chain graph that specifically accounts for  the presence of (deterministic) chains of links. For simplicity, 
we ignore boundary effects that should only give rise to subdominant terms.

We consider a loop between two sites separated by a height $z$ (of course, $z<L$ where $L$ is the overall height of the chains). We study 
loops that take advantage of the chain structure, namely with one part of the loop formed by the shortest segment of length $z$ along 
the chain and the other part formed by a wandering path joining the two sites through sites of intermediate height belonging to different 
chains, all on the same side of the chain to which the two sites belong (we keep a representation of parallel vertical chains as in Fig.~2 
of the main  text). Other types of loops are either longer or more unlikely.
The typical length $\ell_{\rm w}(z)$ of this wandering path can be estimated by considering the random graph obtained by a vertical 
projection of the chain sites up to height $z$ on the base sites at $z=0$, the latter being now connected through all bonds between 
chains: as each site on the chain is randomly connected to $b$ sites at the level above and to $b$ sites at the level below, this graph 
is therefore a quasi-regular random graph with $L_{x}$ sites of average degree $2b z$ (or rather $2bz-b$, but, as already 
mentioned, the fact that sites at $z=0$ are connected to $b$ other chains instead of $2b$ is irrelevant for large $L_{x}$ and $L$). 
The wandering path in the original randomly connected chain graph is now a loop in the projected graph. By using known results on 
random graphs, namely that the typical loop length scales as $\log( L_{x})/\log (2bz)$ \cite{Ballobas,MezardMontanari}, one then obtains an estimate for the total loop 
length $\ell(z)\equiv z+\ell_{\rm w}(z)$ for two sites separated by a height $z$ as
\begin{equation}
\ell(z)\sim z+ \log(L_{x})/\log(2bz) \,.
\end{equation}
The typical loop length can then be obtained by optimizing the above expression with respect to $z$, which gives 
$z^*\sim \log( L_{x})/[\log(\log L_{x})]^2$ and 
\begin{equation}
\ell^* \sim \frac{\log( L_{x})}{\log(\log L_{x})} \,,
\end{equation} 
in the large $L_x$ limit. This expression is valid provided $L>z^*$, which is satisfied if we choose $L$ of the order of $L_x$ as done in the main text. (Note that if $z^*>L$, one obtains {$\ell^*\sim (\log L_{x})/\log L$ that also diverges with $L_x$.) 

\begin{figure}
\begin{center}
\includegraphics[width=4.41cm,trim=8 2 5 5,clip]{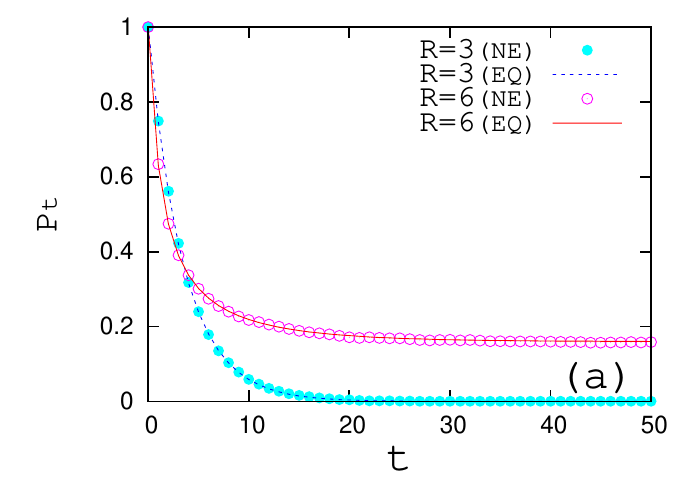}\includegraphics[width=4.41cm,trim=8 2 5 5,clip]{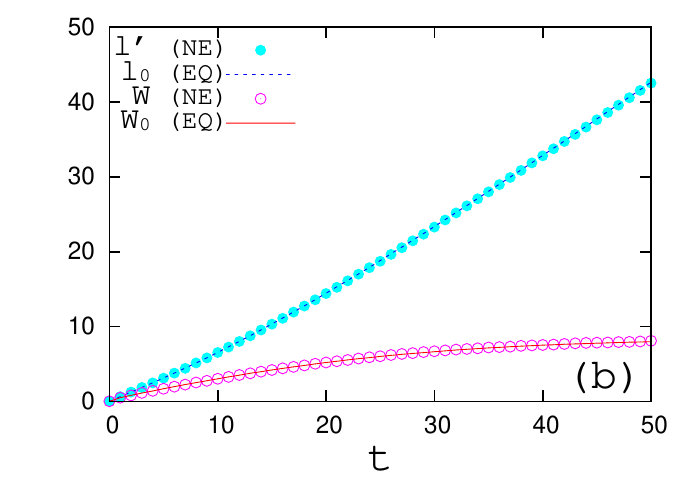}
\caption{Comparison between the solution of the recursion equations (EQ)
and numerical simulation results (NE) on the randomly connected chains with $b=2$  and $L_{x}=10^5$ for one realization 
of the random fields. In all cases, one observes an excellent agreement. (a) $P_t$ obtained from Eq.~(2) and 
computed numerically ($H=2$). 
(b) $l_0(t)$, $W_0(t)$ obtained from Eqs.~(11)-(13) and 
$l'\equiv\langle l_p(t)\rangle$, $W(t,L_{x})$ ($R=6$, $H=2$). }
\label{comp}
\end{center}
\end{figure}

From the above estimate, one can conclude that loops in the randomly connected chain graph
are irrelevant in the limit $L_{x}\rightarrow\infty$, which is what one needs to describe it as a graph with a local tree structure and 
derive recursion equations for the interface-growth model in the thermodynamic limit.

In addition, we have checked that the outcome of the numerical simulations on large randomly connected chain graphs coincide 
with the solution of the recursion equations obtained in the main text by using the tree structure. This is illustrated in Fig.~\ref{comp} 
and confirms the above argument concerning the irrelevance of loops in the 
randomly connected chains in the thermodynamic limit  ($L_{x}\rightarrow \infty$).


\begin{thebibliography}{0}
\bibitem{note1} For convenience, we now define the front as the ensemble of up spins just below the foremost interfacial down spins. This change has no incidence on the results.
\end{thebibliography}
\end{document}